This file contains the uuencoded (figures.tar.gz.uue) ps-figures
attached to the TeX-file below. After separating one should uudecode,
uncopress with gzip and then uncompress with tar typing in the
DOS prompt tar -xvf <file.tar>.

\documentstyle[12pt,fleqn]{article}
\begin{document}
\newcommand{\be}{\begin{equation}}
\newcommand{\ee}{\end{equation}}
\newcommand{\ds}{\displaystyle}
\newcommand{\sss}{\scriptscriptstyle}
\renewcommand{\theequation}{\arabic{section}.\arabic{equation}}
\mbox{ }\hfill{\normalsize ITP-94-30E}\\
\mbox{ }\hfill{\normalsize July 1994}\\
\begin{center}
{\Large \bf Deuteron Electromagnetic Form Factors in the\\[0.5cm]
Transitional Region Between Nucleon-Meson\\[0.5cm] and Quark-Gluon
Pictures}\\
\vspace{1cm}
{\large A.~P.~Kobushkin and A.~I.~Syamtomov}\\[.5cm]
{\large \it N.N.Bogolyubov Institute for Theoretical Physics, \\
National Academy of Sciences of Ukraine, Kiev 143, Ukraine
}
\end{center}
\date{}
\vspace{.5cm}
\begin{abstract}
Experimental observables of the elastic $eD$-scattering in the
region of intermediate energies are discussed. We offer the
analysis of the available experimental data, which
reproduces the results of the calculations with popular
$NN$-potentials at low energies ($Q^2\ll 1(GeV/c)^2$), but,
at the same time, provides the right asymptotic behavior of the
deuteron e.m. form factors, following from the quark counting rules,
at high energies ($Q^2\gg 1(GeV/c)^2$). The numerical analysis
developed allows to make certain estimations of the characteristic
energy scale, at what the consideration of quark-gluon degrees of
freedom in the deuteron becomes essential.
\end{abstract}
\newpage

\section{Introduction}
\setcounter{equation}{0}
It was already demonstrated [1-3], that the quark substructure
of the deuteron should manifest itself in the elastic $ed$-scattering
at high $Q^2$. Nevertheless, up to now the question of the energy scale
at what quark-gluon degrees of freedom
become defrozen in simplest nuclear systems has been remaining a
subject  of great interest (see e.g. Ref. \cite{Kob91}).
It seems that the so called quark counting rules \cite{Ferrar}
applied to the elastic $ed$-scattering at high $Q^2$ give the most
natural way for estimating of such scale.
While QCD predicts \cite{Vainshtein,Gross} that the individual asymptotic
behavior of deuteron form factors will differ significantly
from that predicted by the conventional $NN$-potential models, the
predicted asymptotic behavior of the cross section $\frac{\ds
d\sigma}{\ds dt}\sim {\ds 1\over \ds t^{11}}f({t/s})$ (where $s$ and
$t$ are the conventional Mandelstam variables) can be also reproduced
within the "classical" (nucleon-meson) picture \cite{Woloshin}.
Therefore, it is a topical matter to look for another observables of
the elastic $ed$-scattering which could be sensitive to the
quark-gluon structure of the deuteron.

The aim of the present paper is to study the behavior of the elastic
$ed$\--\-scat\-tering
in intermediate region between nucleon-meson and quark-gluon pictures and
to estimate a $Q^2$-scale, where QCD consideration becomes valid. We
show that the QCD asymptotics sets in the helicity-flip transition
amplitudes at $Q^2$ of order of few $(GeV/c)^2$.
The outline of the paper is as follows. In Sec.2 we summarize the general
expressions for observables of the elastic $ed$-scattering as well as
expressions of the deuteron e.m. form factors in terms of helicity transition
amplitudes in the infinite momentum frame. The QCD predictions for the
helicity transition amplitudes at high
$Q^2$ are discussed in Sec.3. In Sec.4 we
analyze the experimental data for the elastic $ed$-scattering ($A(Q^2)$ and
$B(Q^2)$ structure functions, the tensor analyzing power  $T_{20}$),
using phenomenological parametrization, which reproduces the results
of the nucleon-meson calculations at $Q^{2}\ll 1(GeV/c)^{2}$ and reduces
to the asymptotic behavior predicted by the perturbative QCD at
$Q^{2}\gg 1(GeV/c)^{2}$. It is assumed that the region where $Q^2$ are
of the order of few $(GeV/c)^2$ is the region where the
nucleon-meson calculations as well as pure QCD methods are not
applicable.  Predictions for the e.m. form factors of the deuteron and
$T_{20}$ at $Q^{2}$ of the order of few $(GeV/c)^2$ are done.
Conclusions are given in Sec.5.

\section{Basic relations}
\setcounter{equation}{0}
In the lab. frame the cross section of the elastic $ed$-scattering (when
the particles are unpolarized) is given by the Rosenbluth
formula
\be
{\ds d\sigma \over d\Omega}=\left({\ds d\sigma \over \ds
d\Omega}\right)_{Mott}\left[A\left(Q^2\right)+B\left(Q^2\right)\tan
^{2}\left({\ds \theta \over \ds 2}\right)\right],
\label{2}
\ee
where $\left({\ds d\sigma \over \ds d\Omega}\right)_{Mott}$ is the Mott cross
section, $\theta $ is the electron scattering angle; $A\left(Q^2\right)$
and $B\left(Q^2\right)$ -- are the deuteron structure functions, which,
in turn,
are expressed via the charge, $G_C$, magnetic, $G_M$, and qudrupole, $G_Q$,
deuteron form factors:
\be
A = G_{C}^{2}+{\ds 2\over \ds 3}\eta G_{M}^{2}+{\ds 8\over \ds 9}\eta
^{2}G_{Q}^{2},\ \ \
B = {\ds 4\over \ds 3}\eta (1+\eta )G_{M}^{2},
\label{3}
\ee
where $\eta =Q^2/4M^2$ and $M$ is the deuteron mass.
To separate the charge and qudrupole form factors one has, in addition
to $A$ and $B$ structure functions, to measure polarization observables
of the process, e.g. the tensor analyzing power
\begin{eqnarray}
\hspace{-0.5cm} \lefteqn{T_{20}(Q^{2},\theta )=-\frac{{8\over 9}
\eta ^{2}G^{2}_{Q}+ {8\over 3}\eta
G_{C}G_{Q}+{2\over 3}\eta G^{2}_{M}\left[{1\over 2}+(1+\eta )\tan
^{2}\left(\theta \over 2\right)\right]}{\sqrt{2}
\left[A\left(Q^2\right)+B\left(Q^2\right)\tan ^{2}\left({\theta \over
2}\right)\right]}.}
\label{15}
\end{eqnarray}
The normalization of the form factors entering (\ref{3}) and (\ref{15}) is
chosen to be the following:
$G_{C}(0)=1,\ G_{M}(0)={\ds 2M\over \ds e}\mu _{D}$, $G_{Q}(0)={\ds
M^{2}\over \ds e}Q_{D}$, where $\mu _{D}$ and $Q_{D}$ are the deuteron
magnetic and qudrupole moments, respectively. In the infinite momentum
frame, defined as in Ref. \cite{Drell}, the form factors could be expressed
in terms of
the helicity transition amplitudes \cite{Hiller}
$J_{\lambda^{\prime }\lambda}^{\mu }=<p^{\prime }\ \lambda^{\prime
}|j^{\mu}|p\ \lambda >$:
\begin{eqnarray}
G_{C} & = & \frac {\ds 1}{\ds {2p^{+}(2\eta +1)}}\left[(1-{\ds 2\over
\ds 3}\eta )J^{+}_{00}+{\ds 8\over \ds 3}\sqrt{2\eta
}J^{+}_{+0}+{\ds 2\over \ds 3}(2\eta -1)J^{+}_{+-}\right], \nonumber \\
G_{M} & = & \frac {\ds 1}{\ds {2p^{+}(2\eta +1)}}\left[2J^{+}_{00}+\frac
{\ds 2(2\eta -1)}{\ds \sqrt{2\eta }}J^{+}_{+0}-2J^{+}_{+-}\right],
\label{9} \\
G_{Q} & = & \frac {\ds 1}{\ds {2p^{+}(2\eta +1)}}\left[-J^{+}_{00}+\sqrt{{\ds
2\over \ds \eta}}J^{+}_{+0}-{\ds {\eta +1}\over \ds \eta
}J^{+}_{+-}\right], \nonumber
\end{eqnarray}
where $j^{\mu}$ is the e.m. current,
$|p\ \lambda >$ stands for the deuteron state with the momentum $p$ and
helicity $\lambda $, $J_{\lambda^{\prime }\lambda }^{+}\equiv
J_{\lambda^{\prime}\lambda }^{0}+J_{\lambda^{\prime}\lambda }^{3}$ and
$p^{+}\equiv p^{0}+p^{3}$.
\section{Asymptotic behavior}
\label{asymptotic}
\setcounter{equation}{0}
Recently, Brodsky and Hiller \cite{Hiller} have mentioned
that in the analysis of the high energy elastic $ed$-cross section
at least two momentum scales must be distinguished. The first one is
given by the QCD scale \mbox {$\Lambda_{QCD}\approx200MeV/c$} and
determines the perturbative QCD regime. In particular, according
to the arguments of Ref. \cite{Gross}, this scale controls a high $Q^{2}$
suppression of the helicity-flip transition amplitudes
\begin{eqnarray}
\lefteqn{J^{+}_{+0}\approx a\left(\frac{\ds \Lambda _{QCD}}{\ds
\sqrt{Q^{2}}}\right)J^{+}_{00},\ \ \
J^{+}_{+-}\approx b\left(\frac{\ds \Lambda _{QCD}}{\ds
\sqrt{Q^{2}}}\right)^{2}J^{+}_{00},}
\label{11}
\end{eqnarray}
where $a$ and $b$ are some constants.

The second scale is purely kinematical and defined by the deuteron mass $M$.
It was argued \cite{Hiller,Coester} that in the region
\be
Q^{2}\gg 2M\Lambda _{QCD}\approx 0.8(GeV/c)^{2}
\label{12}
\ee
the helicity-conserving transition amplitude $J_{00}^{+}$ dominates, so
the quark con\-tent of the deuteron could reveal itself already in the
experimentally accessible region.

From (\ref{9}) and the QCD-motivated relations (\ref{11}) it follows that
the charge form factor $G_{C}$ should have the lowest leading fall-off
degree, while the form factors $G_{M}$ and $G_{Q}$ are suppressed by a
factor $(Q^{2})^{-1}$. At the same time at high $Q^{2}$ the meson-nucleon
approach (see e.g. Ref. \cite{Gari}) predicts that $G_{C}\propto G_{M}$
and $G_{Q}$ is suppressed by a factor $(Q^{2})^{-1}$. Thus QCD and the
classical nuclear physics give the different predictions for high-$Q^2$
behavior of such quantities as the ratio $B/A$ and $T_{20}$.

It was assumed \cite{Hiller,Coester} that in (\ref{9})
the helicity-flip transition amplitudes $J_{+0}^{+}$ and
$J_{+-}^{+}$ could be omitted for the kinematical region
(\ref{12}). In this case
$J_{00}^{+}$ cancels in the expressions for $T_{20}$ and $B/A$.
However, the calculated behavior of the last one contradicts the
experimental data.

In Ref. \cite{Syamt} it was mentioned that the helicity-flip matrix
element $J^{+}_{+0}$ cannot be neglected in the magnetic form factor.
Moreover, it was demonstrated that $J^{+}_{+0}$ matrix element strongly
affects the behavior of the magnetic form factor at $Q^2$ of a few
$(GeV/c)^2$ and provides a neat parametrization for $B/A$ ratio. On the one
hand, this kinematical region cannot be considered as a region of
pure nucleon-meson physics; on the second hand, it cannot be considered
as a region of pure perturbative QCD as well. In the next section we
shell study
the deuteron e.m. form factors in the framework of phenomenological
model, which reproduces the results of calculations with realistic
NN-potential at low $Q^2$, and provides the behavior predicted by
perturbative QCD at high $Q^2$.
\section{Smooth connection to low-$Q^2$ region and numerical analysis}
\setcounter{equation}{0}
Following the idea of the reduced nuclear amplitudes in QCD
\cite{BrodChert,BrodHill} we define the reduced helicity transition
amplitudes $g^{+}_{00}$, $g^{+}_{+0}$ and $g^{+}_{+-}$, rewritting
(\ref{9}) in the following way
\begin{eqnarray}
G_{C} & = & \frac {\ds G^{2}(Q^{2})}{\ds {(2\eta +1)}}\left[(1-{\ds 2\over
\ds 3}\eta )g^{+}_{00}+{\ds 8\over \ds 3}\sqrt{2\eta
}g^{+}_{+0}+{\ds 2\over \ds 3}(2\eta -1)g^{+}_{+-}\right], \nonumber \\
G_{M} & = & \frac {\ds G^{2}(Q^{2})}{\ds {(2\eta
+1)}}\left[2g^{+}_{00}+ \frac {\ds 2(2\eta -1)}{\ds \sqrt{2\eta
}}g^{+}_{+0}-2g^{+}_{+-}\right],
\label{17} \\
G_{Q} & = & \frac {\ds
G^{2}(Q^{2})}{\ds {(2\eta +1)}}\left[-g^{+}_{00}+ \sqrt{{\ds 2\over
\ds \eta}}g^{+}_{+0}-{\ds {\eta +1}\over \ds \eta }g^{+}_{+-}\right],
\nonumber
\end{eqnarray}
where $G(Q^{2})$ -- is a dipole form factor
$G(Q^{2})=\left(1+{\ds Q^{2}\over \ds \delta ^{2}}\right)^{-2}$, %
 $\delta $ -- is some parameter of order of the nucleon mass.  For
$G_{M}$ and $G_{Q}$ to be finite at $Q^2\rightarrow 0$, the reduced
helicity transition amplitudes should
obey: $g_{00}^{+}\sim O(1),\ g_{+0}^{+}\sim O(Q),\ g_{+-}^{+}\sim O(Q^{2})$,
which justifies the following parametrization:
\begin{eqnarray}
g_{00}^{+}=\sum_{i=1}^{n} {\ds a_{i}\over \ds
{\alpha _{i}^{2}+Q^{2}}
},\nonumber \\
g_{+0}^{+}=Q\sum_{i=1}^{n} {\ds b_{i}\over \ds
{\beta _{i}^{2}+Q^{2}} },\ \ \
g_{+-}^{+}=Q^{2}\sum_{i=1}^{n} {\ds c_{i}\over \ds {\gamma _{i}^{2}
+Q^{2}}}
\label{19}
\end{eqnarray}
with $\left\{a_{i}, \alpha _{i}\right\},\ \left\{b_{i}, \beta _{i}\right\},\
\left\{c_{i}, \gamma _{i}\right\}$ being fitting parameters.
From the quark counting rules \cite{Gross} it follows that the fall-off
behavior of these amplitudes at high $Q^{2}$'s is
$$
g_{00}^{+}\sim Q^{-2},\ \ \ \ \
g_{+0}^{+}\sim Q^{-3},\ \ \ \ \
g_{+-}^{+}\sim Q^{-4},
$$
which, together with the requirement of correct static normalization
(Sec. 2), impose the set of restrictions on
$\left\{a_{i}\right\},\ \left\{b_{i}\right\},\ \left\{c_{i}\right\}$
\begin{eqnarray}
\lefteqn{\sum_{i=1}^{n} {\ds a_{i}\over \ds {\alpha _{i}^{2}}}=1,} \nonumber \\
\lefteqn{\sum_{i=1}^{n} b_{i}=0,\ \ \ \sum_{i=1}^{n} {\ds b_{i}\over \ds
{\beta _{i}^{2}}}={\ds {2-\mu _{D}}\over \ds 2\sqrt{2}M},} \label{20} \\
\lefteqn{\sum_{i=1}^{n} c_{i}=0,\ \ \ \sum_{i=1}^{n} c_{i}\gamma _{i}^{2}=0,
\ \ \
\sum_{i=1}^{n} {\ds c_{i}\over \ds {\gamma _{i}^{2}}}={\ds {1-\mu _{D}-Q_{D}}
\over \ds 4M^{2}}.} \nonumber
\end{eqnarray}

The "masses" $\left\{\alpha _{i}\right\},\ \left\{\beta _{i}\right\}$ and
$\left\{\gamma _{i}\right\}$ define a nonperturbative part of reduced
amplitudes. In our calculations we used the following sequence for each
group of these parameters:
\be
\alpha _{n}^{2}=2M\mu^{(\alpha )},\ \ \alpha _{i}^{2}=\alpha _{1}^{2}+
\frac{\ds {\alpha _{n}^{2}-\alpha _{1}^{2}}}{\ds {n-1}}(i-1),\ \
i=1,\ldots ,n \nonumber
\ee
(similarly, for $\beta _{i}$ and $\gamma _{i}$), where $\mu^{(\alpha)}$,
$\mu^{(\beta)}$ and $\mu^{(\gamma)}$ have the dimension of energy and, in
accordance with (\ref{12}), are to be of order of $\Lambda_{QCD}$.
Results of numerical calculations (n=4, parameters appear in Table 1)
for $A$, $B$ and $T_{20}$ at $\theta =0^{o}$ are shown on Figs.1-3.
Experimental data are taken from Refs. \cite{Bosted,The,Garcon}. Figs. 4
and 5 display the behavior of $G_C$ and $G_Q$. Experimental points are from
the analysis of Ref. \cite{Garcon} The dashed-dot lines correspond to the
asymptotic QCD behavior of the reduced matrix elements:
\begin{equation}
g_{00}^{+(\infty)}\approx \frac{\sum_{i=1}^{n}a_{i}}{Q^{2}}, \ \ \
g_{+0}^{+(\infty)}\approx -\frac{\sum_{i=1}^{n}b_{i}\beta_{i}^{2}}{Q^{3}}, \ \ \
g_{+-}^{+(\infty)}\approx \frac{\sum_{i=1}^{n}c_{i}\gamma_{i}^{4}}{Q^{4}}.
\label{asymp}
\end{equation}
The reduced helicity transition amplitudes $g_{00}^{+}$, $g_{+0}^{+}$ and
$g_{+-}^{+}$ appear in Figs. 6-8.

As was mentioned before, at the region (\ref{12}) the ratios
$J_{00}^{+}/J_{+0}^{+}$ and $J_{+0}^{+}/J_{+-}^{+}$ should be controlled
by the QCD scale parameter $\Lambda_{QCD}$ only. Figs. 9 and 10 demonstrate
that this QCD prediction works well for $J_{+0}^{+}/J_{+-}^{+}$, while
for the ratio $J_{00}^{+}/J_{+0}^{+}$ the QCD asymptotics starts at somewhat
high $Q^2$.
  Estimation of the characteristic energy scale $Q_{QCD}$ at which one
may expect the asymptotic behavior of the observables of the elastic
$ed$-scattering may be determined from
$$
\mbox{max}\left(\frac{\int_{\sss Q_{QCD}}^{\infty}\left|g_{00}^{+(\infty)}-
g_{00}^{+}\right|\ dQ}{\int_{\sss Q_{QCD}}^{\infty}\left|g_{00}^{+}\right|\ dQ},\
\frac{\int_{\sss Q_{QCD}}^{\infty}\left|g_{+0}^{+(\infty)}-
g_{+0}^{+}\right|\ dQ}{\int_{\sss Q_{QCD}}^{\infty}\left|g_{+0}^{+}\right|\ dQ},
\right.
$$
$$
\hskip 2.in
\left. \frac{\int_{\sss Q_{QCD}}^{\infty}\left|g_{+-}^{+(\infty)}-
g_{+-}^{+}\right|\ dQ}{\int_{\sss Q_{QCD}}^{\infty}\left|g_{+-}^{+}\right|\ dQ}
\right)\leq \varepsilon,
$$
where $\varepsilon$ stands for experimental data accuracy.
Using the best-fit parameters (see Table~1) and taking different values
of ${\bf \varepsilon}$ we made estimations (see. Table~2) of $Q_{QCD}$,
based on all experimental data available.

\vspace{1.0cm}

\begin{center}
\begin{tabular}{||c|c|c|c|c||}         \hline
\multicolumn{5}{|c|}{Table 1} \\ \hline\hline
$\backslash i$ & 1    &  2   &   3   &   4  \\ \hline
$a_{i}\ fm^{-2}$        & $2.4818$ & $-10.850$
 & $6.4416$  & see (\ref{20})  \\ \hline
$b_{i}\ fm^{-1}$        & $-1.7654$          & $6.7874$
 & see (\ref{20}) & see (\ref{20})  \\ \hline
$c_{i}$        & $-0.053830$ & see (\ref{20})  & see (\ref{20})
 & see (\ref{20})  \\ \hline
\multicolumn{3}{||c|}{$\alpha _{1}^{2}=1.8591\ fm^{-2}$} &
\multicolumn{2}{l||}{$\mu^{(\alpha )}=0.58327\ GeV/c$} \\ \hline
\multicolumn{3}{||c|}{$\beta_{1}^2=19.586\ fm^{-2}$} &
\multicolumn{2}{l||}{$\mu^{(\beta  )}=0.1\ GeV/c$} \\ \hline
\multicolumn{3}{||c|}{$\gamma_{1}^2=1.0203\ fm^{-2}$} &
\multicolumn{2}{l||}{$\mu^{(\gamma )}=0.17338\ GeV/c$} \\ \hline
\multicolumn{5}{||c||}{$\delta =0.89852\ GeV/c$}\\ \hline
\end{tabular} \\
\vskip 1cm

\begin{tabular}{||c|c||}         \hline
\multicolumn{2}{|c|}{Table 2} \\ \hline\hline
${\bf \varepsilon}$ & $Q_{QCD}$  \\ \hline
 10\%      & $2.92\ GeV/c$  \\ \hline
 7\%       & $3.44\ GeV/c$  \\ \hline
 5\%       & $4.04\ GeV/c$  \\ \hline
 2\%       & $6.28\ GeV/c$  \\ \hline
 1\%       & $8.84\ GeV/c$  \\ \hline
\end{tabular} \\
\end{center}
\section*{Conclusions}
In the present paper we have studied the possible description
of the existing experimental data of the elastic $ed$-scattering
based on the idea of reduced transition amplitudes, the behavior of
which at high $Q^2$ is fixed by the quark counting rules. At low
$Q^2$ our parametrization reproduces the results of the standard
nucleon-meson picture.

The above analysis shows that QCD could strongly affect the behavior
of the deuteron e.m. form factors when $Q^2$ is of order of $10(GeV/c)^2$.
This value of the transferred
momentum scale is mainly defined by the fact that the QCD asymptotics
of the helicity non-flip transition amplitude $J_{00}^{+}$ starts rather
late. Therefore, at intermediate energy the observables independent
of $J_{00}^{+}$ should be
sufficiently sensitive to the quark structure of the deuteron. The
existing experimental data for the ratio
$J_{+0}^{+}/J_{+-}^{+}$ are close to the asymptotic prediction (\ref{11})
at $Q^2\approx 1(GeV/c)^2$. Nevertheless, all experimental points are also in
satisfactory correspondence with conventional $NN$-potential models.
The essential discrepancy between what is predicted from QCD and and/or
nucleon-meson picture starts at $Q^2\approx 4(GeV/c)^2$. QCD predicts
that from $Q^2\approx 2(GeV/c)^2$ the ratio $J_{+0}^{+}/J_{+-}^{+}$ will
tend monotonically to its asymptotic behavior
$(J_{+0}^{+}/J_{+-}^{+})_{asymp}=4.51\times 10^{-2}(Q/\Lambda_{QCD})$.
From the other hand, from the $NN$-potential models it follows that
from $Q^2\approx 4(GeV/c)^2$ this should be an oscillating function.
\vskip 0.5cm

\section*{Acknowledgments}
\vskip 0.2cm

\noindent
The authors express their thank to C.~Carlson and F.~Gross for useful
discussions.
\newpage

\begin{center}
{\bf Figure captions}
\end{center}

\parbox[t]{1in}{Figure 1.}
\parbox[t]{4.5in}{Structure function $A(Q^2)$ of the elastic $ed$-scattering
in the approach of the reduced transition amplitudes (solid line). The
dashed line corresponds to the calculations with Paris potential.}
\vskip 0.5cm

\parbox[t]{1in}{Figure 2.}
\parbox[t]{4.5in}{Structure function $B(Q^2)$. (For notations see the
caption of Fig.1).}
\vskip 0.5cm

\parbox[t]{1in}{Figure 3.}
\parbox[t]{4.5in}{Tensor analyzing power $T_{20}$. (For notations see the
caption of Fig.1).}
\vskip 0.5cm

\parbox[t]{1in}{Figure 4.}
\parbox[t]{4.5in}{Charge form factor of the deuteron $G_{C}$ in the approach
of the reduced transition amplitudes (solid line). The dashed-dot line
stand for QCD asymptotics (see the text for explanation).}
\vskip 0.5cm

\parbox[t]{1in}{Figure 5.}
\parbox[t]{4.5in}{Quadrupole form factor of the deuteron $G_{Q}$. (For
notations see the caption of Fig.4).}
\vskip 0.5cm

\parbox[t]{1in}{Figure 6.}
\parbox[t]{4.5in}{The $g_{00}^{+}$ reduced transition amplitude (solid line).
The dashed-dot line stands for its pure asymptotical residue (4.11).}
\vskip 0.5cm

\parbox[t]{1in}{Figure 7.}
\parbox[t]{4.5in}{The $g_{+0}^{+}$ reduced transition amplitude. (For
notations see the caption of Fig.6).}
\vskip 0.5cm

\parbox[t]{1in}{Figure 8.}
\parbox[t]{4.5in}{The $g_{+-}^{+}$ reduced transition amplitude. (For
notations see the caption of Fig.6).}
\vskip 0.5cm

\parbox[t]{1in}{Figure 9.}
\parbox[t]{4.5in}{The ratio of helicity transition amplitudes $J_{00}^{+}$
and $J_{+0}^{+}$ calculated within the current approximation (solid line).
The dashed-dot line corresponds to the asymptotics (4.11). (The full
circles are to distinguish the negative values of the experimental data).}
\vskip 0.5cm

\parbox[t]{1in}{Figure 10.}
\parbox[t]{4.5in}{The ratio of the helicity-flip transition amplitudes
$J_{+0}^{+}$ and $J_{+-}^{+}$ corresponding to the current approach
(solid line) and to the approach with Paris potential (dashed line).
The dashed-dot line stands for the asymptotics (4.11).}

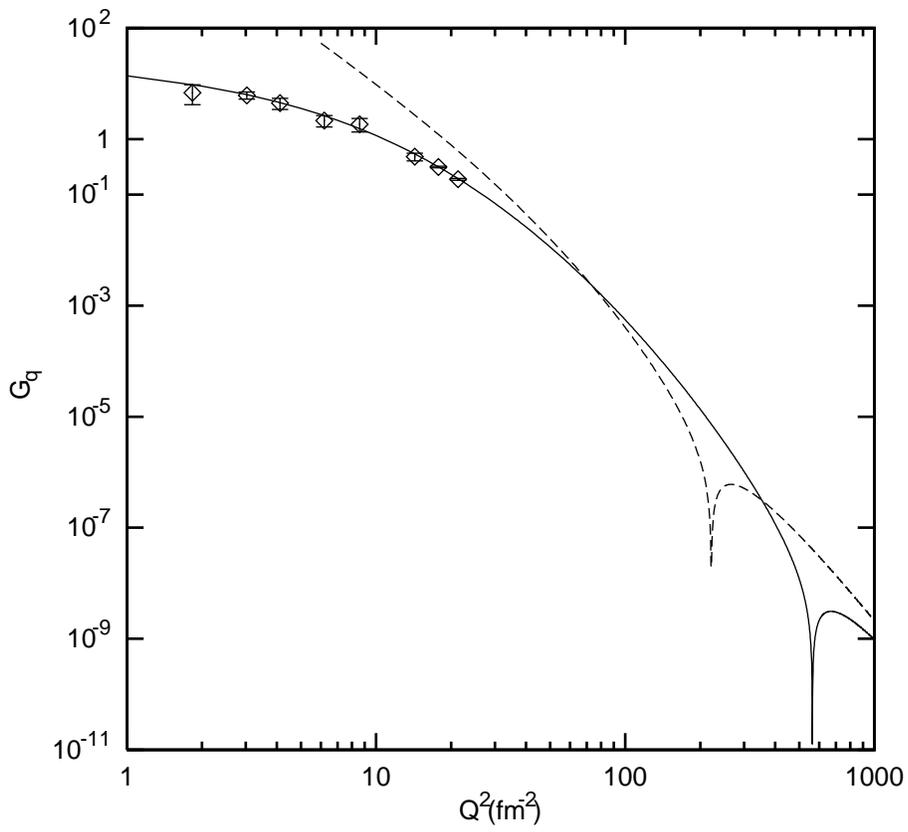
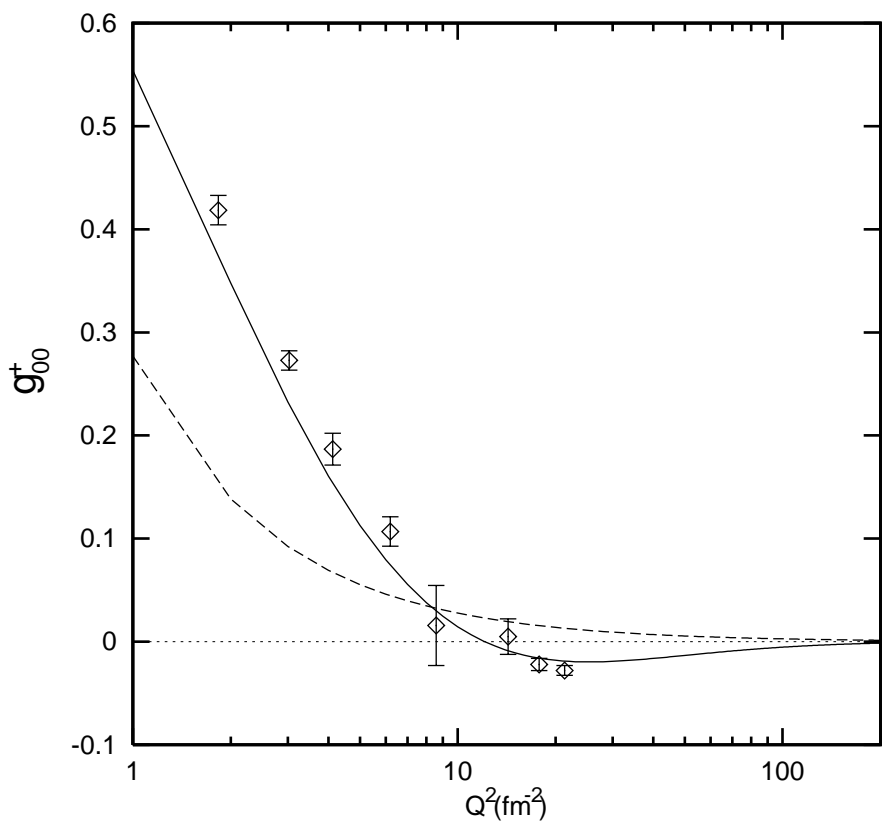
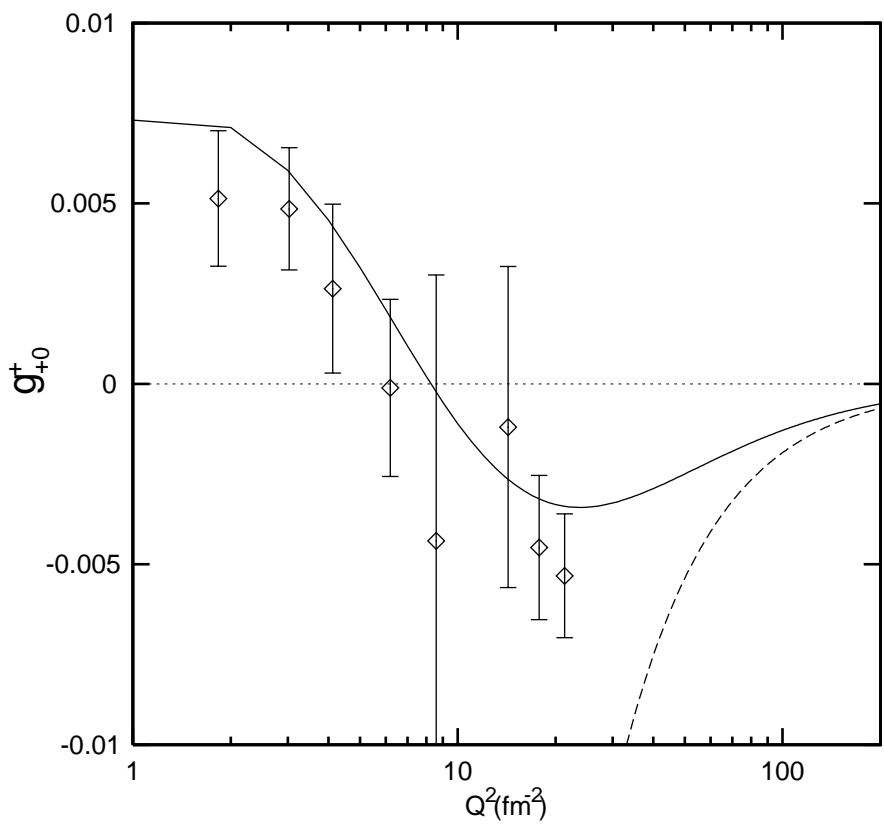
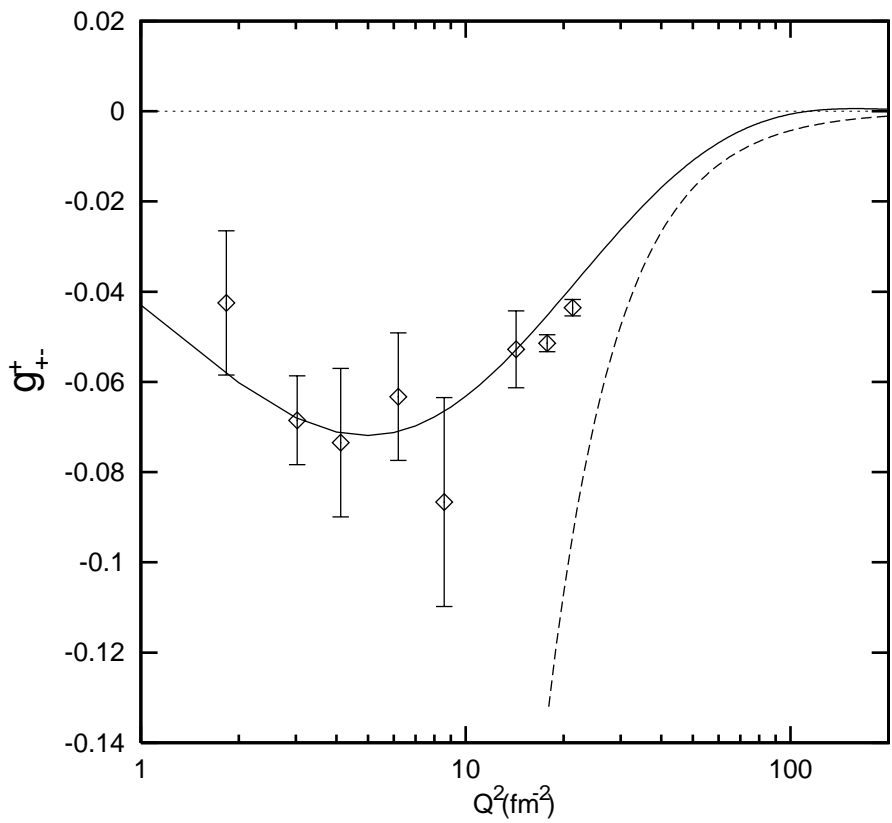
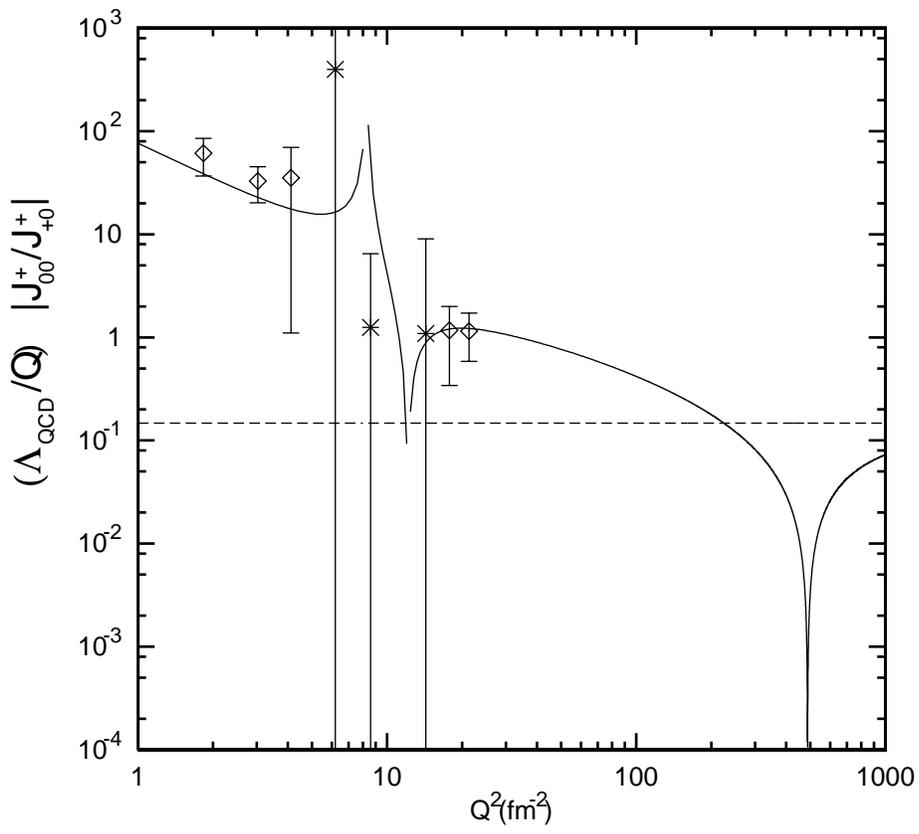
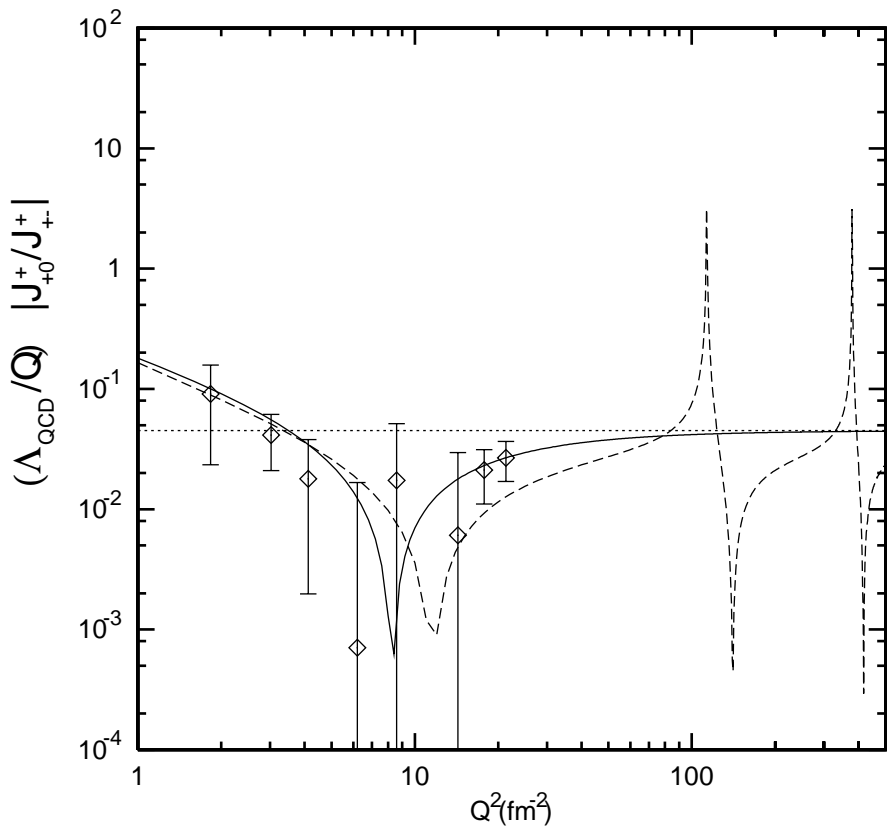
\end{document}